\documentclass[prl,twocolumn,showpacs]{revtex4}
\usepackage{graphicx}
\usepackage{indentfirst}
\usepackage{epsfig}
\usepackage{subfigure}
\usepackage{amsmath}
\usepackage{amssymb}
\def\be{\begin{eqnarray}}
\def\ee{\end{eqnarray}}

\begin{document}
\title{Dipole emission and coherent transport in random media I}
\author{M. Donaire$^{1}$}
\email{manuel.donaire@uam.es}
\address{$^{1}$Departamento de F\'{\i}sica de la Materia Condensada, Universidad Aut\'{o}noma de
Madrid, E-28049 Madrid, Spain.}
\begin{abstract}
This is the first of a series of papers devoted to develop a
microscopical approach to the  dipole emission process and its
relation to coherent transport in random media. In this Letter, we
deduce general expressions for the decay rate of spontaneous
emitters and the power emission of induced dipoles
 embedded in homogenous dielectric media. We derive formulae which
 apply generically to virtual cavities and, in the continuum approximation, to small real cavities.
\end{abstract}
%
 \maketitle
\indent It is a general issue in physics the characterization of a
system by means of its coherent transport properties and the study of the decay of unstable
 local states. In the quite general case that
 the constituents of the medium couple  to each other through dipole-dipole interactions, the decay
 of an excited particle takes place through radiative and non-radiative emission.
  In particular, it is known that the spontaneous emission rate, $\Gamma$,
 in a dielectric medium depends on the interaction of the emitter with the environment \cite{Purcell}. This is so because the
 surrounding medium determines the number of channels into which the excited particle can decay. That is, the local
 density of states (LDOS). It is in this sense that the net effect of the
 random medium is to renormalize the vacuum as seen by the emitter.
  On the other hand, LDOS and $\Gamma$
 depend, upon additional properties of the emitter, on the statistical parameters which determine the
 coherent transport features of the medium. That is, on the electric susceptibility $\bar{\chi}$, the refraction index
  and the mean free path. As a matter of fact, null transmittance
  and inhibition of spontaneous emission are expected to occur in
  photonic band gap materials \cite{JohnI}. The
  understanding of life-times in random medium is relevant in
  the context of fluorescence biological imaging \cite{Suhling}
  and nano-antennas \cite{Antenas}. On the other hand, understanding of unconventional coherent transport properties is essential in engineering metamaterials for electromagnetic  and acoustic waves \cite{PendryLiu}.
In this Letter we deduce general formulae for the spontaneous and
induced dipole power emission in a homogeneous random medium
characterized by its electrical susceptibility. Longitudinal and
transverse components are differentiated. It paves the way for the
deduction of the relation between LDOS and $\Gamma$ with coherent
transport parameters.\\
\indent The power $W^{\mu}_{\omega}$ emitted in the process of
spontaneous decay of a point dipole from an excited
 state $\Psi(\omega)$ is directly  proportional to its decay rate $\Gamma^{\mu}_{\omega}$.
 The relation is given by
$\Gamma^{\mu}_{\omega}=\frac{4}{\omega\hbar\epsilon_{0}}W^{\mu}_{\omega}$,
where
\begin{eqnarray}\label{laprimera}
W^{\mu}_{\omega}&=&\frac{\omega\epsilon_{0}}{2}\Im{\{\vec{\mu}\cdot\vec{E}^{*}_{exc}\}}=
\frac{\omega^{3}}{2c^{2}}\Im{\{\vec{\mu}\cdot\bar{\mathcal{G}}^{*}_{\omega}(\vec{r},\vec{r})\cdot\vec{\mu}^{*}\}}
\nonumber\\&=&
-\frac{\omega^{3}}{6c^{2}}|\mu|^{2}\Im{\Bigl\{\textrm{Tr}\{\bar{\mathcal{G}}_{\omega}(\vec{r},\vec{r})\}\Bigr\}}.
\end{eqnarray}
In the above equation, Tr is the trace operator, $\mu$ is the dipole transition amplitude,
$\omega$ is the transition frequency and
$\bar{\mathcal{G}}_{\omega}(\vec{r},\vec{r})$ is the propagator of
the field emitted by the dipole in the process back to itself. We
denote 2-rank tensors with overlines. In general, $\mu$, $\omega$
and $\bar{\mathcal{G}}_{\omega}$ all depend
on the interaction between the emitter and
the host medium. On the other hand, LDOS is proportional
to $\Im{\Bigl\{\textrm{Tr}\{\bar{\mathcal{G}}_{\omega}(\vec{r},\vec{r})\}\Bigr\}}$
 and so are $\Gamma^{\mu}_{\omega}$ and $W^{\mu}_{\omega}$ \cite{Wylie}.\\
\indent Next, consider that the emitter consists of the dipole
induced by a fixed exciting field
$\vec{E}_{exc}(\vec{r})=\vec{E}^{\omega}_{0}(\vec{r})$ on a
scatterer of radius $a$ and dielectric contrast
$\epsilon_{e}(\omega)$. $\omega$ is assumed far from any resonance frequency of the emitter.
Smallness implies $a\ll k_{0}^{-1}$, $k_{0}=\omega/c$ being the
bare wave number. Thus, $\vec{E}^{\omega}_{0}(\vec{r})$ is uniform
within the scatterer. If the emitter is one of the host scatterers
in a homogeneous and isotropic random medium, the emitted power
reads
\begin{eqnarray}\label{laleche}
W^{\alpha}_{\omega}&=&
\frac{\omega\epsilon_{0}^{2}}{2}\Im{}\Bigl\{\int\textrm{d}^{3}r\:\chi^{\omega}_{e}\:\Theta(r-a)\int\textrm{d}^{3}r'
\textrm{d}^{3}r''
\bar{\mathbb{G}}_{\omega}(\vec{r},\vec{r}')\nonumber\\&\cdot&[\bar{G}^{(0)}]^{-1}(\vec{r}',\vec{r}'')
\cdot\vec{E}^{\omega}_{0}(\vec{r}'')\cdot
\vec{E}^{\omega*}_{0}(\vec{r})\Bigr\},
\end{eqnarray}
where $\chi^{\omega}_{e}=(\epsilon_{e}(\omega)-1)$ --not to be
confused with the susceptibility of the random medium-- and
\begin{equation}\label{b3}
\bar{\mathbb{G}}_{\omega}(\vec{r})=\bar{G}^{(0)}(\vec{r})
\sum_{m=0}^{\infty}\Bigl[-k_{0}^{2}\chi^{\omega}_{e}\int\Theta(v-a)\bar{\mathcal{G}}_{\omega}(v)\textrm{d}^{3}v\Bigr]^{m}
\end{equation}
$\bar{G}^{(0)}(\vec{r})$ being the tensor propagator of the
electric field in free space. In spatial space representation, it
consists of a Coulombian field propagator
$\bar{G}_{Co.}^{(0)}(r)=\Bigl[\frac{1}{k_{0}^{2}}
\vec{\nabla}\otimes\vec{\nabla}\Bigr]\Bigl(\frac{-1}{4\pi\:r}\Bigr)$
 plus a radiation field propagator, $\bar{G}_{rad.}^{(0)}(r)=\frac{e^{i\omega r}}{-4\pi
r}\mathbb{I}+\bigl[\frac{1}{k_{0}^{2}}\vec{\nabla}\otimes\vec{\nabla}\bigr]\frac{e^{i\omega
r}-1}{-4\pi r}$. In Eq.(\ref{laleche}),
$\bar{\mathbb{G}}_{\omega}(\vec{r},\vec{r}')$ is the propagator of
the field emitted at a point $\vec{r}'$ inside the particle back
to another point $\vec{r}$ also within the particle. Notice that,
because the emitter is polarizable in this case, the wave comes
back and forth
 infinite times with propagator $\bar{\mathcal{G}}_{\omega}$. Hence, series Eq.(\ref{b3}). In the above
 equation, for $a\ll k_{0}^{-1}$,  the electric field is
 nearly uniform and we can approximate
$\int\Theta(r-a)\bar{\mathcal{G}}_{\omega}(r)\simeq\frac{4\pi}{3}a^{3}\bar{\mathcal{G}}_{\omega}(0)$.
This expression is formally correct. However, the
perturbative expansion of $\bar{\mathcal{G}}_{\omega}(0)$ contains
singularities hidden in both the electrostatic and radiative parts of
$\bar{G}^{(0)}$. The later is related to resonance frequencies of the emitter polarizability \cite{RMPdeVries} and so, negligible in the present case. The former is regularized by considering the finite size of the emitter, Lim$\{\int\textrm{d}^{3}r\:\Theta(r-a)\bar{G}^{(0)}(r)\}=\frac{1}{3k_{0}^{2}}\mathbb{I}$
as $a\rightarrow0$. One way to avoid carrying this limit in
further calculations  consists of dressing up the single particle
susceptibility with all the 'in-vacuum' electrostatic corrections
at once. That is, by defining
$\tilde{\chi}^{\omega}_{e}\equiv\frac{3}{\epsilon_{e}+2}\chi^{\omega}_{e}$
and the electrostatic polarizability $\alpha_{0}\equiv 4\pi
a^{3}\frac{\epsilon_{e}-1}{\epsilon_{e}+2}$. On the other hand,
the part free of singularities can be written as
\begin{figure}[h]
\includegraphics[height=3.1cm,width=8.6cm,clip]{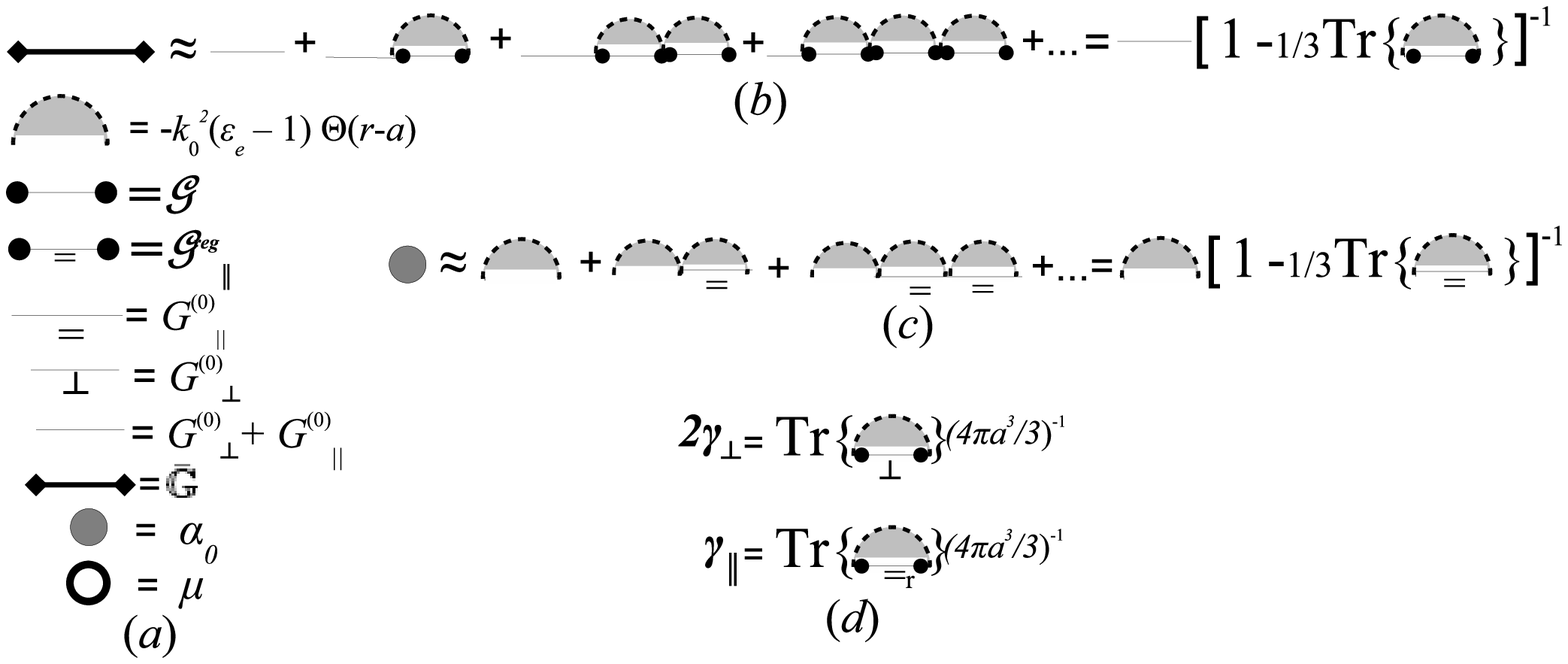}
\caption{($a$) Feynman's rules. ($b$) Diagrammatic representation
of Eq.(\ref{b3}). ($c$) Diagrammatic representation of the
dressing up of $\chi_{e}^{\omega}$ leading to $\alpha_{0}$.
Approximation symbols denote that the field within the emitter is
taken uniform. ($d$) Diagrammatic representation of
Eq.(\ref{b5}).}\label{fig21}
\end{figure}
\begin{eqnarray}\label{b5}
\bar{\mathcal{G}}_{\omega}^{reg}(0)&=&\frac{1}{3}\Bigl[\int\frac{\textrm{d}^{3}k}{(2\pi)^{3}}2\mathcal{G}^{reg}_{\perp}(k)\:
+\:\int\frac{\textrm{d}^{3}k}{(2\pi)^{3}}\mathcal{G}^{reg}_{\parallel}(k)\Bigr]\mathbb{I}\nonumber\\
&\equiv&\frac{1}{3}\Bigl[2\gamma_{\perp}+\gamma_{\parallel}\Bigl]\mathbb{I},
\end{eqnarray}
where $\mathcal{G}^{reg}_{\perp,\parallel}(k)$ are
singularity-free and we have defined the
$\gamma_{\perp,\parallel}$ factors in the last row. In
Eq.(\ref{b5}), the scripts  $\parallel$ and $\perp$ denote, in
Fourier space, the tensor components along and transverse to the
propagation direction respectively. Note that longitudinal and
transverse modes couple to each other in the series Eq.(\ref{b3})
as scattering takes place off the dipole surface. With the above
definitions we can rewrite Eq.(\ref{laleche}) in terms of
electrostatically renormalized operators -- compare with
\cite{RMPdeVries,LagvanTig},
\begin{eqnarray}\label{laleche2}
W^{\alpha}_{\omega}&=&
\frac{\omega\epsilon_{0}^{2}}{2}\Im{}\Bigl\{\int\textrm{d}^{3}r\:\tilde{\chi}^{\omega}_{e}\:\Theta(r-a)\int\textrm{d}^{3}r'
\textrm{d}^{3}r''
\bar{\tilde{\mathbb{G}}}_{\omega}(\vec{r},\vec{r}')\nonumber\\&\cdot&[\bar{G}^{(0)}]^{-1}(\vec{r}',\vec{r}'')
\cdot\vec{E}^{\omega}_{0}(\vec{r}'')\cdot
\vec{E}^{\omega*}_{0}(\vec{r})\Bigr\},
\end{eqnarray}
where
\begin{equation}\label{b4}
\bar{\tilde{\mathbb{G}}}_{\omega}(\vec{r},\vec{r}')\equiv\bar{G}^{(0)}(\vec{r},\vec{r}')
\sum_{m=0}^{\infty}(-k_{0}^{2}\alpha_{0})^{m}3^{-m}\Bigl(2\gamma_{\perp}+\gamma_{\parallel}\Bigr)^{m}.\nonumber
\end{equation}
The power emitted/absorbed by the induced dipole according to
Eq.(\ref{laleche2}) is
\begin{eqnarray}
W^{\alpha}_{\omega}&=&
\frac{\omega\epsilon_{0}^{2}}{2}\Im{}\Bigl\{\frac{\alpha_{0}}{1+\frac{1}{3}k_{0}^{2}\alpha_{0}[2\gamma_{\perp}+\gamma_{\parallel}]}
\Bigr\}|E^{\omega}_{0}|^{2}\label{la1}\\&=&\frac{-\omega^{3}\epsilon_{0}^{2}}{6c^{2}}\Bigl\{\frac{|\alpha_{0}|^{2}}
{|1+\frac{1}{3}k_{0}^{2}\alpha_{0}[2\gamma_{\perp}+\gamma_{\parallel}]|^{2}}\Im{\{2\gamma_{\perp}+\gamma_{\parallel}\}}\label{la2}\\
&-&
\frac{\Im{\{\alpha_{0}\}}}{|1+\frac{1}{3}k_{0}^{2}\alpha_{0}[2\gamma_{\perp}+\gamma_{\parallel}]|^{2}}\Bigr\}|E^{\omega}_{0}|^{2}\label{la3}.
\end{eqnarray}
The term in Eq.(\ref{la3}) corresponds to the power
 absorbed within the emitter. The term in Eq.(\ref{la2})
corresponds to the power radiated into the medium. The latter
contains contributions of both coherent and incoherent radiation
together with power
absorbed by the host scatterers.\\
\indent Finally, consider the spontaneous emission of a point
dipole like that in Eq.(\ref{laprimera}), but now with an
additional in-vacuum polarizability $\alpha=\alpha_{0}[1-i
k_{0}^{3}\alpha_{0}/(6\pi)]^{-1}$. The situation is analogous to
that of a fluorescent particle placed on top of a host scatterer.
Thus, the dipole moment of the system emitter-host-particle reads
$\vec{p}=\vec{\mu}+\frac{\omega^{2}}{c^{2}}\tilde{\alpha}\bar{\mathcal{G}}_{\omega}(0)\vec{\mu}$,
$\tilde{\alpha}$ being the renormalized polarizbility of the host
particle. The spontaneous field emitted in the decay process gives
rise to an induced dipole moment in the host particle which
modifies the decay rate of the emitter. It is plain from the
perturbative development in Fig.\ref{fig22} that the emitted power
can be written as
\begin{eqnarray}
W^{\alpha,\mu}_{\omega}&=&
\frac{\omega\epsilon_{0}^{2}}{2}\frac{|\mu|^{2}}{\epsilon_{0}^{2}}\Im{}\Bigl\{\alpha_{0}^{-2}\Bigl[\frac{\alpha_{0}}{1+\frac{1}{3}k_{0}^{2}\alpha_{0}[2\gamma_{\perp}+\gamma_{\parallel}]}
-\alpha_{0}\Bigr]\Bigr\}\label{latercera}\nonumber\\&=&\frac{-\omega^{3}}{6c^{2}}\frac{|\mu|^{2}}
{|1+\frac{1}{3}k_{0}^{2}\alpha_{0}[2\gamma_{\perp}+\gamma_{\parallel}]|^{2}}
\Bigl[\Im{\{2\gamma_{\perp}+\gamma_{\parallel}\}}\label{latercera1}\\
&-&\frac{k_{0}^{2}}{3}\Im{\{\alpha_{0}\}}|2\gamma_{\perp}+\gamma_{\parallel}|^{2}\Bigr]\label{latercera2},
\end{eqnarray}
where we recognize again  the power absorbed within the
induced dipole in the last term and the power radiated into the
surrounding medium in the remaining.
\begin{figure}[h]
\includegraphics[height=1.9cm,width=8.8cm,clip]{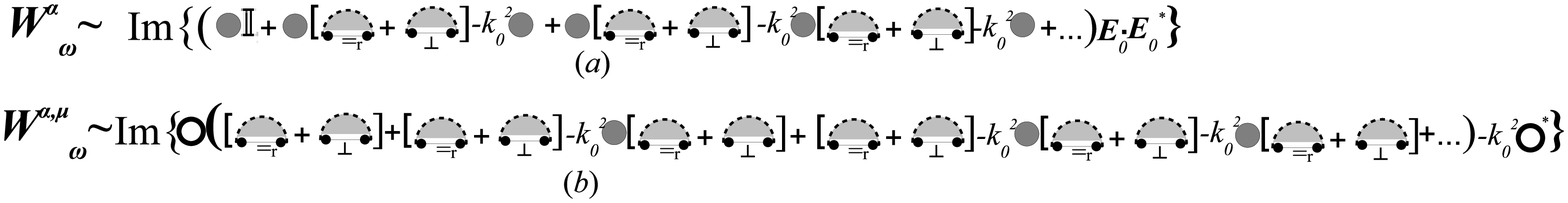}
\caption{($a$) Diagrammatic representation of Eq.(\ref{laleche}).
 ($b$) Diagrammatic representation of Eqs.(\ref{latercera1},\ref{latercera2}).}\label{fig22}
\end{figure}
 \indent Emission in vacuum is obtained by setting
$\bar{\mathcal{G}}_{\omega}(r)=\bar{G}^{(0)}(r)$ in all the
equations above.\\
\indent For a complete description of $W$, $\Gamma$ and LDOS we are just left with the computation of
$\bar{\mathcal{G}_{\omega}}(\vec{r},\vec{r})$ and its trace
components, $2\gamma_{\perp}$ and $\gamma_{\parallel}$ in each case.
 $\bar{\mathcal{G}_{\omega}}(\vec{r},\vec{r})$ is the propagator of the field throughout the bulk from the emitter
back to itself. It includes multiple-scattering processes which are in general spatially correlated. Correlations are  both those among host scatterers themselves and those of host scatterers with the emitter. If the emitter does not perturb the statistical isotropy of the host medium, it
occupies the center of a spherical cavity of radius $R$.
On top of that, the infinite series in Eq.(\ref{b3}) amount to multiple self-polarization processes.
It is convenient to formulate additional  Feynman's rules to describe the cavity and the self-polarization
cycles. Those are given in Fig.\ref{fig23}. A self-polarization cycle carries a factor
$[2\gamma_{\perp}+\gamma_{\parallel}]\frac{1}{3}\mathbb{I}=\int$d$^{3}r\bar{\mathcal{G}}^{reg}_{\omega}(r)\delta^{(3)}(\vec{r})$
in the perturbative expansions for $W$ in place of
$(\frac{4\pi
   a^{3}}{3})^{-1}\int$d$^{3}r\bar{\mathcal{G}}^{reg}_{\omega}(r)\Theta(r-a)$ --see Fig.\ref{fig24}($c,d$).
The cavity gives rise to a negative correlation
function $h_{C}=-\Theta(r-R)$.\\
\indent In addition, we have to describe the field propagation
throughout the bulk. Special attention is to be paid to the
interaction between longitudinal and transverse modes. For
simplicity let us assume the host medium is infinite. Therefore,
no coupling to surface modes needs to be considered.  A bulk
propagator $\bar{G}^{\omega}$ and dielectric and susceptibility
tensors, $\bar{\epsilon}^{\omega}$ and $\bar{\chi}^{\omega}$
 can be unambiguously defined. $\bar{G}^{\omega}$ is the dyadic
  Green's function of the macroscopic
Maxwell equations for the time-mode $\omega$ in the bulk. We drop
the script $\omega$ hereafter in all quantities. $\bar{\chi}$
carries correlation effects and is thus made of
one-particle-irreducible (1PI) multiple-scattering events. In Fourier
space, translation invariance and isotropy allow us to split Dyson
equation for $\bar{G}(k)$ in two uncoupled and mutually orthogonal
scalar algebraic equations,
\begin{eqnarray}
G_{\perp}(k)&=&G_{\perp}^{(0)}(k)\:-\:k_{0}^{2}\:G_{\perp}^{(0)}(k)\:\chi_{\perp}(k)\:G_{\perp}(k),\label{DysonI}\\
G_{\parallel}(k)&=&G_{\parallel}^{(0)}(k)\:-\:k_{0}^{2}\:G_{\parallel}^{(0)}(k)\:\chi_{\parallel}(k)\:G_{\parallel}(k),
\label{DysonII}
\end{eqnarray}
where $\bar{G}_{\perp}^{(0)}(k)=\frac{\Delta(\hat{k})}{k_{0}^{2}-k^{2}}$ and
$\bar{G}_{\parallel}^{(0)}(k)=\frac{\hat{k}\otimes\hat{k}}{k_{0}^{2}}$,
$\hat{k}$ being a unitary vector along the propagation direction
and $\Delta(\hat{k})\equiv I-\hat{k}\otimes\hat{k}$ being the
projective tensor orthogonal to $\hat{k}$. The longitudinal
component is the  propagator of the electrostatic field. Likewise,
the transverse component is the propagator of the
radiation field. In view of
Eqs.(\ref{DysonI},\ref{DysonII}), longitudinal and transverse
coherent photons do not couple to each other as traveling
throughout a random medium as it is the case of photons in free
space. Eqs.(\ref{DysonI},\ref{DysonII}) can be solved
independently yielding the renormalized propagator functions for
the coherent --macroscopic-- electric field,
$\bar{G}_{\perp}(k)=\frac{\Delta(\hat{k})}{k_{0}^{2}\epsilon_{\perp}(k)-k^{2}}$,
$\bar{G}_{\parallel}(k)=\frac{\hat{k}\otimes\hat{k}}{k_{0}^{2}\epsilon_{\parallel}(k)}$.
However, a more detailed examen shows that longitudinal and
transverse bare photons --i.e., normal modes of $\bar{G}^{(0)}$--
do couple necessarily when they propagate in a random medium  and
experience multiple scattering processes. In
Eqs.(\ref{DysonI},\ref{DysonII}), longitudinal and transverse bare
photons enter both $\chi_{\perp}(k)$ and $\chi_{\parallel}(k)$ by
means of the spatial correlations among scatterers. In particular,
coupling between longitudinal and transverse modes shows up at
the emitter cavity.
\begin{figure}[h]
\includegraphics[height=1.7cm,width=6.8cm,clip]{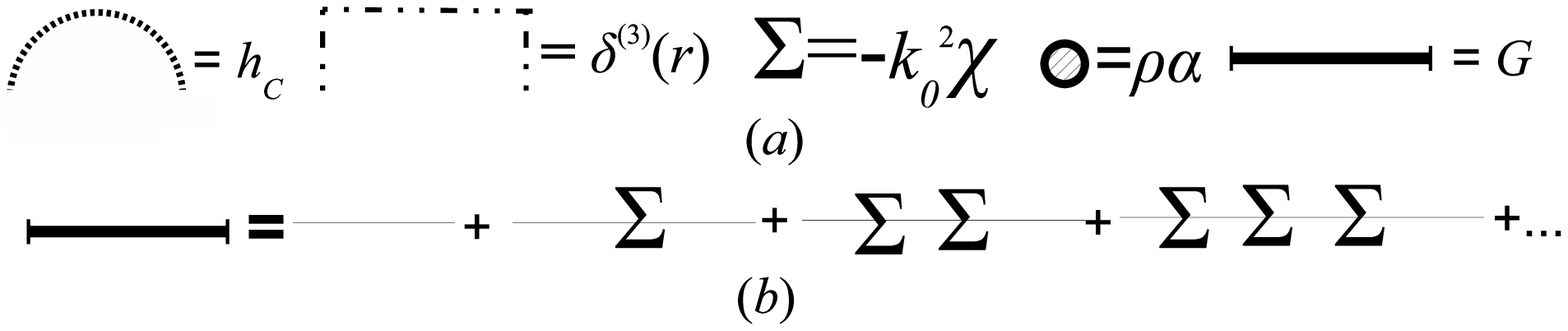}
\caption{($a$) Feynman's rules. ($b$) Diagrammatic representation
of the bulk propagator $\bar{G}$.}\label{fig23}
\end{figure}
We define the cavity factors
\begin{eqnarray}
C_{\perp}(k)&=&\frac{1}{2}\int\textrm{d}^{3}r\:e^{i\vec{k}\cdot\vec{r}}h_{C}(r)\textrm{Tr}
\{\bar{G}^{(0)}(r)[\bar{I}-\hat{k}\otimes\hat{k}]\}\nonumber\\&=&
\frac{1}{2}\int\frac{\textrm{d}^{3}k'}{(2\pi)^{3}}
h_{C}(|\vec{k}'-\vec{k}|)\Bigl[G_{\perp}^{(0)}(k')\nonumber\\&+&G_{\perp}^{(0)}(k')\cos^{2}{\theta}\:+\:G_{\parallel}^{(0)}(k')\sin^{2}{\theta}\Bigr],
\label{Xiperp}\\
C_{\parallel}(k)&=&\int\textrm{d}^{3}r\:e^{i\vec{k}\cdot\vec{r}}h_{C}(r)\textrm{Tr}
\{\bar{G}^{(0)}(r)[\hat{k}\otimes\hat{k}]\}\nonumber\\&=&\int\frac{\textrm{d}^{3}k'}{(2\pi)^{3}}
h_{C}(|\vec{k}'-\vec{k}|)\nonumber\\&\times&\Bigl[G_{\parallel}^{(0)}(k')\cos^{2}{\theta}\:+\:G_{\perp}^{(0)}(k')\sin^{2}{\theta}\Bigr],
\label{Xiparall}
\end{eqnarray}
where $\cos{\theta}\equiv \hat{k}\cdot\hat{k}'$.\\
\indent Consider first the emitter as either a spontaneous or induced dipole
equivalent in all to the rest of host scatterers or as a
fluorescent particle on top of a host scatterer. Therefore, it
correlates to the surrounding as any host particle would do. The exclusion
volume around the emitter is thus a virtual cavity as it does not perturb the medium at all. In each self-polarization cycle,
the emitted/polarizing field experiences in the bulk
multiple-scattering processes which are correlated to the emitter/receiver.
A typical process of these  is depicted in Fig.\ref{fig24}($b$).
There, only 2-point correlation functions, $h(r)$, are used for
simplicity. Note that, because the emitter and the receiver
coincide, the correlations of any intermediate scattering process
to one or the other extreme of the diagram are equivalent. This
allows to attribute all the correlations to the emitter on the
left, which amounts to the factor $\chi/\rho\alpha$
--Figs.\ref{fig24}($c,d$). Therefore, we have
\begin{eqnarray}
2\gamma_{\perp}&=&
\int\frac{\textrm{d}^{3}k}{(2\pi)^{3}}\Bigl[\frac{2\chi_{\perp}(k)/(\rho\alpha)}{k_{0}^{2}[1+\chi_{\perp}(k)]-k^{2}}\Bigr]-2\Re{\{\gamma^{(0)}_{\perp}\}},
\label{LDOSIper}\\
\gamma_{\parallel}&=&\int\frac{\textrm{d}^{3}k}{(2\pi)^{3}}\:\Bigl[\frac{1}{\rho\alpha}\frac{\chi_{\parallel}(k)}{k_{0}^{2}[1+\chi_{\parallel}(k)]}-\frac{1}{k_{0}^{2}}\Bigr],\label{LDOSIparal}
\end{eqnarray}
where $\rho$ is the density of host scatterers. The factor $2$ in front of $\gamma_{\perp}$ stands for the two
transverse polarizations while there is only one longitudinal. The term $-2\Re{\{\gamma^{(0)}_{\perp}\}}$
accounts for the singularity of $\bar{G}_{rad}^{(0)}(0)$.
This concludes the computation of the  spontaneous emission of an emitter on top of a host
scatterer as a function of the susceptibility tensor $\bar{\chi}$
of the host medium and the scatterer polarizability $\alpha_{0}$. We
can split up the power emitted and the associated decay rates into
longitudinal and transverse components, $2\Gamma_{\perp}$ and
$\Gamma_{\parallel}$, as it derives from
Eq.(\ref{latercera1}) (compare with \cite{LagvanTig}). It is plain by substitution of
$\bar{\chi}(k)\approx\rho\alpha-(\rho\alpha)^{2}k_{0}^{2}\bar{C}$
in Eqs.(\ref{DysonI},\ref{DysonII}) that, for any $h_{C}(r)\neq
const$, longitudinal and transverse photons do couple at the
cavity surface. This is at the root of the contribution of
longitudinal modes to the total decay rate, even in absence of
absorbtion (see \cite{MeII}). In particular, for
$R\ll k_{0}^{-1}$, the only contribution to
Eqs.(\ref{Xiperp},\ref{Xiparall}) comes from longitudinal photons
--i.e. electrostatics, yielding
$C_{\parallel,\perp}(k)=\frac{-1}{3k_{0}^{2}}$. This is recognized
as the Lorentz-Lorenz (LL) cavity factor \cite{Lorentz}.\\
\indent Next, let us consider that the emitter is either an
impurity itself or seats on top of a polarizable impurity within
an arbitrary cavity. That breaks manifestly translation
invariance. Thus, the cavity is real and it is not strictly
possible to define a Dyson propagator.
The reason being that, in
any multiple scattering process, beside the translation-invariant
correlation functions joining host scatterers, each scatterer $i$
is correlated to the emitter through the 2-point correlation
function $g_{C}(r_{i})\equiv1+h_{C}(r_{i})$. A first approximation
to go around this problem can be made if $R\gg\xi$, $\xi$ being
the typical correlation length between host scatterers. In such a
case we can consider that, for any 1PI diagram in
$\bar{\chi}(\vec{r}_{1},\vec{r}_{2})$, the emitter gets correlated
to any of the host scatterers therein by simple convolution of
$\bar{\chi}(\vec{r}_{1},\vec{r}_{2})$ with $g_{C}(r_{1})$. This
yields the series expansion depicted in Fig.\ref{fig25}($b$) in
which wave propagation still depends on the
emitter location. A further approximation consists of considering
that, for $k_{0}R\ll1$ the empty cavity do not affect the wave propagation but in
its entrance and departure from the host medium. That is, both wave and
emitter see a continuum beyond $r\sim R$. This approximation is
closer to the Onsager--B\"{o}ttcher (OB) \cite{Onsager} approach
as reformulated in \cite{BulloughHynne}  (see also the quantum approach in \cite{GL}). If $G$ is the bulk
propagator as defined in absence of cavity, the $\gamma$ functions take the form --see
Fig.\ref{fig25}($c$),
\begin{eqnarray}
2\gamma_{\perp}&\simeq&-i\frac{k_{0}}{2\pi}-2k_{0}^{2}\int\frac{\textrm{d}^{3}k}{(2\pi)^3}
[C_{\perp}+G_{\perp}^{(0)}]\chi_{\perp}G_{\perp}^{(0)}(k)\nonumber\\&+&
2k_{0}^{4}\int\frac{\textrm{d}^{3}k}{(2\pi)^3}[C_{\perp}+G^{(0)}_{\perp}]^{2}
G_{\perp}\chi_{\perp}^{2}(k),\label{OB2a}\\
\gamma_{\parallel}&\simeq&-k_{0}^{2}\int\frac{\textrm{d}^{3}k}{(2\pi)^3}
[C_{\parallel}+G_{\parallel}^{(0)}]\chi_{\parallel}G_{\parallel}^{(0)}(k)\nonumber\\&+&
k_{0}^{4}\int\frac{\textrm{d}^{3}k}{(2\pi)^3}[C_{\parallel}+G^{(0)}_{\parallel}]^{2}
G_{\parallel}\chi_{\parallel}^{2}(k),\label{OB2b}
\end{eqnarray}
\begin{figure}[h]
\includegraphics[height=1.8cm,width=6.8cm,clip]{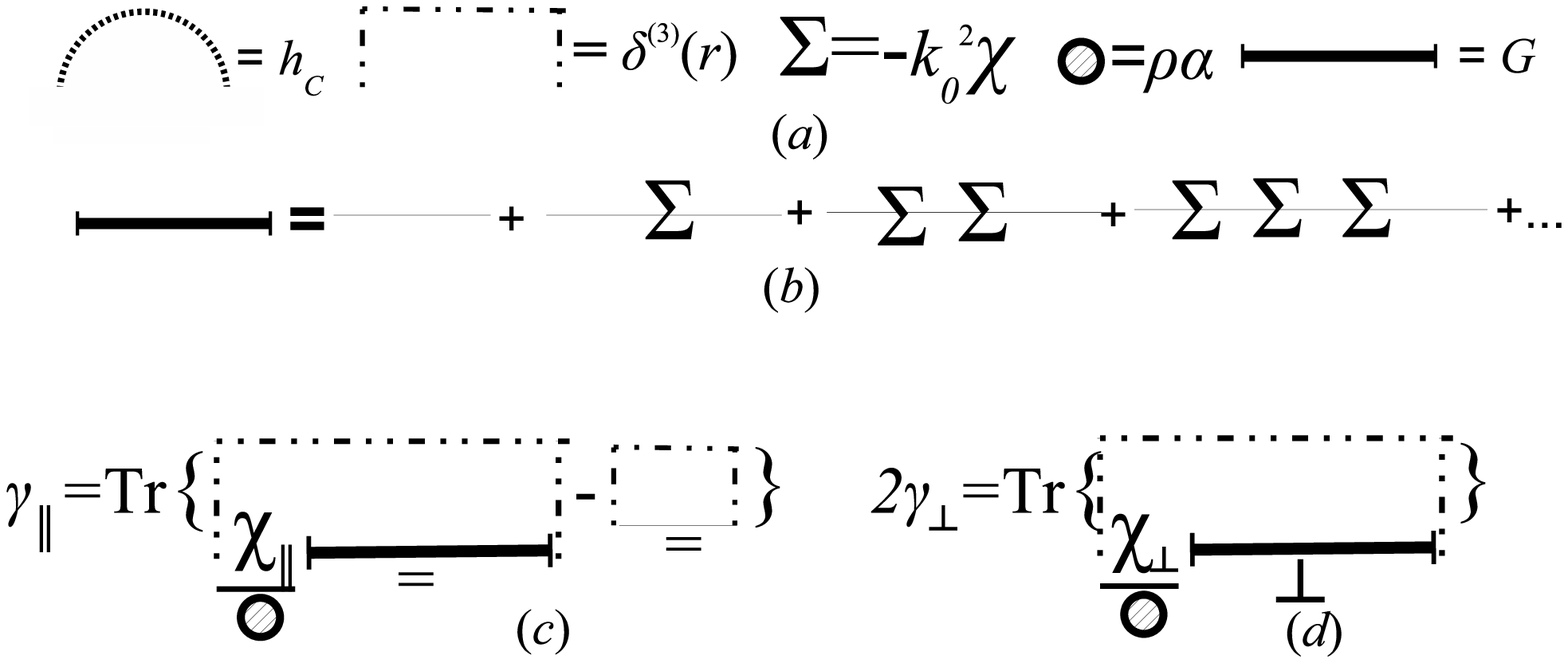}
\caption{($a$) 2-point correlation function $h(r)$ and associated
interaction vertex. ($b$) Diagrammatic representation of the
equivalence between multiple-scattering processes amounting to
$\bar{\mathcal{G}}$. ($c$),($d$) Alternative representation to
Fig.\ref{fig21}($d$). }\label{fig24}
\end{figure}
\begin{figure}[h]
\includegraphics[height=2.1cm,width=7.7cm,clip]{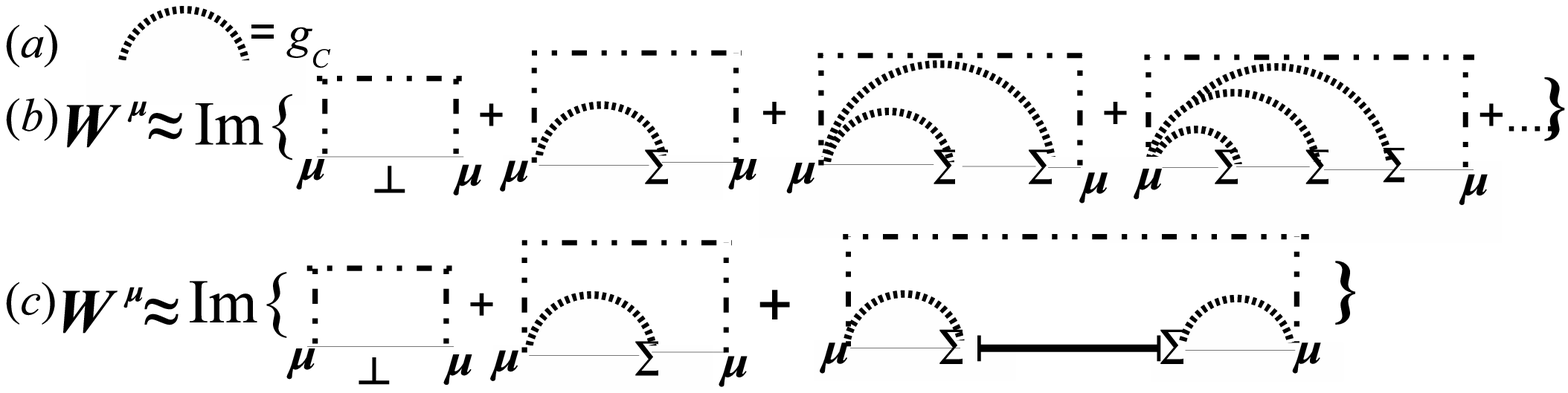}
\caption{($a$) 2-point correlation function $g_{c}(r)$. ($b$)
First approximation of $W^{\mu}$. ($c$) Diagrammatic
representation of the approximate formulas in
Eqs.(\ref{OB2a},\ref{OB2b}).}\label{fig25}
\end{figure}
where an analogous equivalence to that illustrated in
Fig.\ref{fig24}($b$) has been used.\\
\indent In summary, we have derived general expressions for the
longitudinal and transverse components of the power emission and the decay rate of induced dipoles and spontaneous emitters respectively in function of the electrical susceptibility of
the host medium. These are Eqs.(\ref{laprimera};\ref{la2},\ref{la3};\ref{latercera1},\ref{latercera2}), depending on the emitter nature, together with the appropriate $\gamma$-factors. Those are, Eqs.(\ref{LDOSIper},\ref{LDOSIparal}) for a source seated in a virtual cavity. They are exact and their diagrammatic representation is that of
Fig.\ref{fig22}($b$) --up to constant prefactors. For a source seated in a real cavity drilled in a continuous medium,
the $\gamma$-factors are in Eqs.(\ref{OB2a},\ref{OB2b}). They are approximate and their diagrammatic representation
is that of Fig.\ref{fig25}($c$).
%
\indent We thank S.Albaladejo, R.Carminati and J.J.Saenz for
fruitful discussions and suggestions. This work has been supported by Microseres-CM  and the EU
Integrated Project "Molecular Imaging" (LSHG-CT-2003-503259).


\begin{thebibliography}{99}
\bibitem{Purcell} E.M. Purcell,  Phys. Rev. {\bf 69}, 681 (1946).
\bibitem{JohnI} E. Yablonovitch,  Phys. Rev. Lett. {\bf 58}, 2059
(1987); S. John,  Phys. Rev. Lett. {\bf 58}, 2486 (1987); S. John
and T. Quang,  Phys. Rev. A {\bf 50}, 1764 (1994).
\bibitem{Suhling} K. Suhling, P.M.
W. French and D. Phillips, Photochem. Photobiol. Sci. {\bf 4}, 13
(2005).
\bibitem{Antenas} V.V. Protasencko, A.C. Gallagher, Nano. Lett. {\bf 4}, 1329 (2004);
J.N. Farahani, D.W. Pohl, H.J. Eisler and B. Hecht, Phys. Rev.
Lett. {\bf 95}, 017402 (2005); R. Carminati \emph{et al.}, Opt. Com. {\bf 261} (368).
\bibitem{PendryLiu} Z. Liu \emph{et al.}  Science {\bf 289}, 1734 (2000); D.R. Smith, J.B. Pendry and M.C.K. Wiltshire, Science {\bf 305}, 788 (2004).
\bibitem{Wylie} J.M.Wylie, J.E. Sipe, Phys. Rev. A {\bf 30}, 1185 (1984).
\bibitem{RMPdeVries}
P. de Vries, D.V. van Coevorden, A.Lagendijk,
Rev. Mod. Phys. {\bf 70}, 447 (1998).
\bibitem{LagvanTig} A.Lagendijk, B.van Tiggelen, Phys.Rep. {\bf 270}, 143 (1996).
\bibitem{GL}R.Glauber, M.Lewenstein, Phys. Rev. A {\bf 43}, 467 (1991).
\bibitem{Lorentz} H.A.Lorentz, Wiedem. Ann. {\bf 9}, 641 (1880);
 L.Lorenz, Wiedem. Ann. {\bf 11}, 70 (1881).
\bibitem{PRLdeVries}
P.de Vries, A.Lagendijk, Phys.Rev.Lett. {\bf 81}, 1381(1998).
\bibitem{MeII} M.Donaire, e-print arXiv:0811.0373.
\bibitem{Onsager} L.Onsager, J. Am. Chem. Soc. {\bf 58}, 1486
(1936).
\bibitem{BulloughHynne} F. Hynne, R.K. Bullough J. Phys. A {\bf 5}, 1272 (1972); F. Hynne, R.K. Bullough, Phil. Trans. R. Soc. Lond. A {\bf 321}, 305 (1987).
\end{thebibliography}
\end{document}